# CONCAVE AND CONVEX PHOTONIC BARRIERS IN GRADIENT OPTICS

*By*

Alexander B. Shvartsburg[1] and Guillaume Petite[2]

(1) *Central Design Bureau for Unique Instrumentation of the Russian Academy of Sciences*
Butlerov Str. 15, Moscow, Russian Federation
(2) *Laboratoire des Solides Irradiés, UMR 7642, CEA-DSM, CNRS et Ecole Polytechnique*,
F-91128, Palaiseau, France

Abstract :

Propagation and tunneling of light through photonic barriers formed by thin dielectric films with continuous curvilinear distributions of dielectric susceptibility across the film, are considered. Giant heterogeneity-induced dispersion of these films, both convex and concave, and its influence on their reflectivity and transmittivity are visualized by means of exact analytical solutions of Maxwell equations. Depending on the cut-off frequency of the film, governed by the spatial profile of its refractive index, propagation or tunneling of light through such barriers are examined. Subject to the shape of refractive index profile the group velocities of EM waves in these films are shown to be either increased or deccreased as compared with the homogeneous layers; however, these velocities for both propagation and tunneling regimes remain subluminal. The decisive influence of gradient and curvature of photonic barriers on the efficiency of tunneling is examined by means of generalized Fresnel formulae. Saturation of the phase of the wave tunneling through a stack of such films (Hartman effect), is demonstrated. The evanescent modes in lossy barriers and violation of Hartman effect in this case is discussed.

**I - Introduction.**

This paper is devoted to the traveling and tunneling regimes of propagation of electromagnetic waves through thin dielectric layers with continuous spatial distributions of dielectric susceptibility in the direction of propagation $\varepsilon(z)$. This problem has a long history, starting from the first analytic results, obtained by Rayleigh for waves whose velocity inside the medium depends linearly upon the coordinate /1/. Later the linear profile for $\varepsilon(z)$ /2/ as well as an exponential and more general Epstein profiles /3/ were used for the analysis of radio propagation in the ionosphere. Some more complicated distributions were modeled by piece-wise profiles of $\varepsilon(z)$ /4/, described by a WKB approximation /5/ or treated numerically /6/. These researches focused on the propagation of EM waves in heterogeneous media with positive $\varepsilon$, although the tunneling phenomena, arising when $\varepsilon < 0$, were touched sometimes too, e.g., in the case of radiowaves percolation nearby the ionospheric maxima /7/.

The advent of lasers attracted attention upon light tunneling in a series of optoelectronics problems, such as, e.g., the evanescent modes in dielectric waveguides /8/, surface waves on microspheres /9/, Goos – Hanchen effect for optical coatings /10/. A new burst of interest into these phenomena was stimulated by the intriguing perspective of superluminal light propagation through opaque barriers /11/. The experiments in microwave range with "undersized" waveguide /12/ and bi-prism device /13/ as well as the analysis of spatial displacement of the peak of a tunneling pulse /14/ and the direct measurement of photons' tunneling time /15/ were considered by some authors in favor of the concept of superluminal phase time for the tunneling EM waves /16/. However, this concept aroused controversial viewpoints /17/, /18/.

The theoretical background of the aforesaid researches is based on the rectangular model of opaque barrier, pioneered by Gamov for the theory of $\alpha$ – decay as long ago as in 1928 /19/. Another model, combining the well known rectangular and linear barriers (the "trapezoidal" barrier) was developed in

/20/. However, these models, built from the broken straight lines, fail to interpret the salient features of photonic barriers formed by continuous smooth variations of material's transparency.

On the contrary, this paper is intended to bridge the gap between the traveling and evanescent regimes of wave propagation through concave and convex photonic barriers, formed by continuously distributed heterogeneities of $\varepsilon(z)$. These regimes are shown to be determined by strong heterogeneity - induced dispersion ( HID ), dependent upon the finite spatial scales of heterogeneity profile. Such non – local dispersive effects are visualized by means of exact analytical solutions of Maxwell equations for curvilinear photonic barriers, as we showed in /21/, and will be recalled briefly for completeness hereunder (Section II). These solutions, obtained without any suppositions about smallness or slowness of variations of fields or media, visualize the decisive influence of gradient and curvature of photonic barrier on it is reflectivity and transmittivity. The group velocities of energy transfer through heterogeneous films $v_g$ in traveling (Section III) and tunneling (Section IV) regimes, while remaining subluminal, are shown to be either accelerated or delayed as compared with $v_g$ for the homogeneous film.

The tunneling of light wave through an arbitrary amount of thin dielectric nanolayers with concave profile $\varepsilon(z)$ is considered in Section V. The phase shift of the evanescent wave is shown to tend to some constant limit after passage through a stack containing a finite amount of layers. A superluminal phase time, provided by this saturation of phase shift, is found. In section VI, we study the case of lossy films and show that they present specific features.

**II - Heterogeneity – induced dispersion of thin dielectric films.**

We recall here some results exposed in a previous paper /21/ on the antireflection properties of such inhomogeneous films. To visualize the effects of HID let us consider a simple problem of normal incidence of linearly polarized EM wave with components $E_x$ and $H_y$, propagating in the $z$ – direction, incidenting on the interface $z = 0$ of a heterogeneous, non-magnetic and lossless material, described by some continuous distribution of dielectric susceptibility in the transparent region $\varepsilon(z) > 0$, $z >= 0$ in a form

$$\varepsilon(z) = n_0^2 U^2(z) \qquad U\big|_{z=0} = 1 \qquad (1)$$

Here $n_0$ is the value of refractive index of material on the boundary $z = 0$. Expressing the field components $E_x$ and $H_y$ through the vector – potential **A** ( $A_x = \psi$, $A_y = A_z = 0$ ), one can reduce the system of Maxwell equations, related to this geometry, to one equation governing the function $\psi$,

$$\frac{\partial^2 \psi}{\partial z^2} - \frac{n_0^2 U^2(z)}{c^2}\frac{\partial^2 \psi}{\partial t^2} = 0 \qquad (2)$$

We will examine the solution of (2) for the profiles U(z)

$$U(z) = \left(1 + \frac{s_1 z}{L_1} + \frac{s_2 z^2}{L_2^2}\right)^{-1} ; \; s_1 = 0, \pm 1 ; \; s_2 = 0, \pm 1 \qquad (3)$$

containing two spatial scales $L_1$ and $L_2$. In the case of opposite signs of $s_1$ and $s_2$ the profile (3) has either a maximum ($s_1 = -1$, $s_2 = +1$) or a minimum ($s_1 = +1$, $s_2 = -1$) with a value $U_m$

$$U_m = (1+s_1 y^2)^{-1} \qquad y = L_2/2L_1 \tag{4}$$

The scales $L_1$ and $L_2$ are linked in these cases with the thickness $d$ and the values of $U_m$ (4) : $L_1 = d/4y^2$, $L_2 = d/2y$. The solution of the heterogeneous wave equation (2) with the profile $U$ (3) can be written as a spatially non-sinusoidal wave /21/

$$\psi = [U(z)]^{-\frac{1}{2}} \exp[i(q\eta - \omega t)]; \quad \eta = \int_0^z U(z_1) dz_1 \tag{5}$$

$$q = \frac{\omega n_0}{c} N; \quad N^2 = 1 - \frac{\Omega^2}{\omega^2} \tag{6}$$

The characteristic frequencies in (6) are different for concave ($\Omega_1$) and convex ($\Omega_2$) profiles :

$$\Omega_1^2 = \frac{c^2(1+y^2)}{n_0^2 L_2^2}; \qquad \Omega_2^2 = \frac{c^2(y^2-1)}{n_0^2 L_2^2} \tag{7}$$

We note that $\eta$ is proportional to the optical phase path at the position $z$ and that the quantity $n_0 N$ has the meaning of a refractive index. It is remarkable that $N$, as well as the characteristic frequencies $\Omega_1$ and $\Omega_2$ are determined by the spatial scales $L_1$ and $L_2$ only, and not by the characteristic frequencies of the material itself.

The salient features of this heterogeneity – induced dispersion (HID) are:

1. Subject to the interplay of scales $L_1$ and $L_2$ for concave or convex profiles $U(z)$, parameters $\Omega^2$ in (6) can have positive, negative or zero values. Herein, in a case $\Omega^2 > 0$, the factor $N$ resembles the refractive index for a plasma or waveguide with cut – off frequency $\Omega$, dependent upon the profile $\varepsilon(z)$.

The values of $\Omega$ for a nanolayer with thickness $d$ of about 100 nm, refractive index $n_0 \approx 1.8$ and modulation depth of about 25% may be as high as $2.2 \cdot 10^{15}$ rad s$^{-1}$, which relates to a free space wavelength in a near IR range.

2. The wave with frequency, less then cut-off frequency for the concave photonic barrier, is passing through this barrier in a tunneling regime.

3. The variations of wave velocity inside the medium, produced by HID, can exceed the analogous effects of material dispersion by several orders of magnitude.

Electric and magnetic components of the EM field can be found by means of the vector–potential component $A_x = \psi$ (5)

$$E_x = \frac{i\omega C}{c}\psi \qquad (8)$$

$$H_y = \frac{i\omega n_0 C U(z)}{c}\left[1 + \frac{iU_z}{2qU^2}\right]\psi; \quad U_z = \frac{dU}{dz} \qquad (9)$$

where C is an amplitude factor. The solutions (8)–(9) are valid for arbitrary wavelengths and spatial scales of heterogeneity. In the limit $L_2 \to 0$ the profile (3) is reduced to the well known Rayleigh profile which is, thus, a limiting case of the profiles under discussion. Moreover, when both scales $L_1$ and $L_2$ are increasing and, thus, the heterogeneity is vanishing ($\Omega^2 \to 0$, $N \to 1$)), solution (7) is reduced in this limit to a WKB approximation, where these HID effects are dropped. In the particular case, where the heterogeneity scales $L_{1,2}$ are finite, but $\Omega_2 = 0$, this approach reveals the peculiar $U(z)$ profiles, e.g., $U(z) = (1 + z/L)^{-2}$, for which the WKB solutions remain exact /22, 23/.

**III - Group velocities of traveling EM waves in concave and convex photonic barriers.**

The spatial waveforms of the EM field inside the heterogeneous medium are non – sinusoidal and this field is formed due to interference of forward and backward waves; so, the group velocities $v_g$ of these waveforms have to be found by means of energy flux (Pointing vector) **P** and energy density $W$ /7/ :

$$\mathbf{v}_g = \frac{\mathbf{P}}{W}; \quad \mathbf{P} = \frac{c}{4\pi}\mathrm{Re}\left[\mathbf{E} \wedge \mathbf{H}^*\right] \tag{10}$$

$$W = \frac{1}{8\pi}\left(\varepsilon|\mathbf{E}|^2 + |\mathbf{H}|^2\right) \tag{11}$$

To stress out the physical phenomena initiated by the heterogeneity, we will consider a film in air, without substrate. The spatial structure of the EM field inside the barrier is formed by the interference of forward wave, passing through the plane z = 0, and backward one, reflected from the plane z = d. Using formulae (8) – (9) one can present these waves in a form

$$E_x = \frac{i\omega C}{c}\left(e^{iq\eta} + Qe^{-iq\eta}\right)\frac{1}{\sqrt{U}}$$

$$H_y = iqC\sqrt{U}\left[\frac{iU_z}{2qU^2}\left(e^{iq\eta} + Qe^{-iq\eta}\right) + e^{iq\eta} - Qe^{-iq\eta}\right] \tag{12}$$

where

$$U_z = -\frac{2U^2}{qL_2}\left(ys_1 + \frac{zs_2}{L_2}\right) \tag{13}$$

For simplicity the time – dependent factor exp($-i\omega t$) is omitted here and below. The dimensionless parameter $Q$ describes the reflectivity of the far boundary $z = d$. Introducing the reflection coefficient of the film $R$, we can write the continuity conditions on the plane $z = 0$ :

$$E_i(1+R) = \frac{i\omega C}{c}(1+Q) \tag{14}$$

$$E_i(1-R) = \frac{i\omega n_0 NC}{c}\left[-\frac{is_1}{2qL_1}(1+Q)+1-Q\right] \tag{15}$$

Here $E_i$ is the electric component of the incidenting wave, so that the amplitude $C$ writes :

$$C = \frac{-icE_i(1+R)}{1+Q} \tag{16}$$

and parameter $Q$ can be found from the continuity conditions on the interface $z = d$ :

$$Q = \frac{-\exp(2iq\eta_0)\left(1-\frac{i}{2}s_1\gamma - n_0 N\right)}{1-\frac{i}{2}s_1\gamma + n_0 N} \qquad \gamma = c/\omega L_1 \tag{17}$$

The value $\eta_0 = \eta(d)$ will be obtained from (5). The field components $E_x$ (12) and $H_y$ (13) can be rewritten using eqs. (16) and (17)

$$E_x = \frac{E_i(1+R)}{(1+Q)\sqrt{U}}\left(e^{iq\eta} + Qe^{-iq\eta}\right) \tag{18}$$

$$H_y = \frac{E_i(1+R)n_0 N}{1+Q}\sqrt{U}\left\{e^{iq\eta}\left[1-\frac{i}{qL_2}\left(s_1 y+\frac{s_2 z}{L_2}\right)\right]-Qe^{-iq\eta}\left[1+\frac{i}{qL_2}\left(s_1 y+\frac{s_2 z}{L_2}\right)\right]\right\} \quad (19)$$

Substitution of (18) – (19) into (10) brings the expression for the EM energy flow $P_z$

$$P_z = \frac{c}{\pi}\frac{|M|^2 (n_0 N)^2 |E_i|^2}{|\Delta|^2}$$

$$|M|^2 = \frac{(1+R)(1+R^*)}{(1+Q)(1+Q^*)}; \quad \Delta = 1+n_0 N - \frac{is_1\gamma}{2}$$

(20)

We thus note that the energy flow $P_z$ is coordinate – independent, while on the contrary, the EM energy density W proves to be coordinate – dependent : substitution of (18) – (19) into (11) permits one to describe this dependence by means of a dimensionless function $\theta_+$ :

$$W = \frac{|M|^2 (n_0 N)^2}{4\pi |\Delta|^2} U\theta_+ \quad (21)$$

with

$$\theta_+ = n_0^2 + \frac{1+\gamma^2/4}{N_+^2}+\left[1+\frac{\gamma^2}{4}+n_0^2 N_+^2\right]\left[1+\frac{1}{(qL_2)^2}\left(y-\frac{z}{L_2}\right)^2\right]+...$$

$$\cos[2q(\eta_0-\eta)]\left\{n_0^2-\frac{1+\gamma^2/4}{N_+^2}+\left[1+\frac{\gamma^2}{4}-n_0^2 N_+^2\right]\left[1-\frac{1}{(qL_2)^2}\left(y-\frac{z}{L_2}\right)^2\right]+\frac{2\gamma n_0 N_+}{qL_2}\left(y-\frac{z}{L_2}\right)\right\}+...(22)$$

$$s_1\sin[2q(\eta_0-\eta)]\left\{-\frac{\gamma n_0}{N_+}-\frac{2}{qL_2}\left[1+\frac{\gamma^2}{4}-n_0^2 N_+^2\right]\left(y-\frac{z}{L_2}\right)+\gamma n_0 N_+\left[1-\frac{1}{(qL_2)^2}\left(y-\frac{z}{L_2}\right)^2\right]\right\}$$

Here the subscripts "+" in $\theta_+$ and $N_+$ indicates that we are considering the case $N^2 > 0$. Finally, making use of (10), we will find the group velocity $v_g$ inside the curvilinear barrier

$$v_g = \frac{4c}{U\theta_+} \qquad (23)$$

It is worth stressing out that this expression for the group velocity is valid for both convex and concave profiles of $\varepsilon(z)$ with $N^2 > 0$.

**IV - Group velocity of evanescent waves in concave barriers.**

If the wave frequency is less than the cut – off frequency, the radiation flux will be transmitted through the film in the tunneling regime. This situation can arise in a transparent film due to concave profile $U(z)$ (eq.3) ($s_1 = +1$, $s_2 = -1$). Introducing the notations

$$p = \frac{\omega}{c}n_0 N_-; \quad N_- = \sqrt{u^2 - 1}; \quad u = \frac{\Omega}{\omega} > 1 \qquad (24)$$

and proceeding as above, one can write the field components for the evanescent waves

$$E_x = \frac{E_i(1+R)}{(1+Q)\sqrt{U}}\left(e^{-p\eta} + Qe^{p\eta}\right) \qquad (25)$$

$$H_y = i\frac{E_i(1+R)n_0 N_-}{1+Q}\sqrt{U}\left\{e^{-p\eta}\left[1 - \frac{1}{pL_2}\left(y + \frac{z}{L_2}\right)\right] - Qe^{p\eta}\left[1 + \frac{1}{pL_2}\left(y - \frac{z}{L_2}\right)\right]\right\} \qquad (26)$$

Parameter Q, connected with the "reflection" of the evanescent wave on the far side of film, is

$$Q = \frac{\exp(-2p\eta_0)\left(n_0 N_- + \frac{\gamma}{2} + i\right)}{n_0 N_- - \frac{\gamma}{2} - i} \qquad (27)$$

It is remarkable, that the presentation of EM field, tunneling through the heterogeneous film, can be found from the relevant formulae for the traveling wave through the following replacements

$$q \to ip; \quad N_+ \to iN_-$$

$$\cos[2q(\eta_0 - \eta)] \to \text{ch}[2p(\eta_0 - \eta)]; \quad \sin[2q(\eta_0 - \eta)] \to i\,\text{sh}[2p(\eta_0 - \eta)] \qquad (28)$$

Thus eq. (12), (13) and (17) are mapped to eq. (25), (26) and (27) respectively. Using this replacement, one can write energy flux $P_z$ and density W in the forms, similar to (20) – (21), with the function

$$\theta_- = n_0^2 \frac{1+\gamma^2/4}{N_-^2} + \left[1 + \frac{\gamma^2}{4} - n_0^2 N_-^2\right]\left[1 - \frac{1}{(pL_2)^2}\left(y - \frac{z}{L_2}\right)^2\right] + \ldots$$

$$\text{ch}[2p(\eta_0 - \eta)]\left\{n_0^2 + \frac{1+\gamma^2/4}{N_-^2} + \left[1 + \frac{\gamma^2}{4} + n_0^2 N_-^2\right]\left[1 + \frac{1}{(pL_2)^2}\left(y - \frac{z}{L_2}\right)^2\right] + \frac{2\gamma n_0 N_-}{pL_2}\left(y - \frac{z}{L_2}\right)\right\} - \ldots \quad (29)$$

$$\text{sh}[2p(\eta_0 - \eta)]\left\{\frac{\gamma n_0}{N_-} + \frac{2}{pL_2}\left[1 + \frac{\gamma^2}{4} + n_0^2 N_-^2\right]\left(y - \frac{z}{L_2}\right) + \gamma n_0 N_-\left[1 + \frac{1}{(pL_2)^2}\left(y - \frac{z}{L_2}\right)^2\right]\right\}$$

Substitution of $\theta_-$ (29) instead of $\theta_+$ into (23) yields the group velocity $v_g$ of the evanescent wave. Transition from the traveling modes ($\omega \geq \Omega$, $N^2 \geq 0$) to the evanescent ones ($\omega \leq \Omega$, $N^2 \leq 0$) is continuous in $N = 0$, both functions $\theta_-$ and $\theta_+$ having in this point identical values

$$\theta_+\big|_{N=0} = \theta_-\big|_{N=0} =$$
$$2\left[1+n_0^2+\frac{n_0^2}{1+y^2}\left(2y-\frac{z}{L_2}\right)^2 - \frac{2(\eta_0-\eta)}{L_2}\left\{\frac{n_0^2}{1+y^2}\left[1+y^2+\left(y-\frac{z}{L_2}\right)^2\right]+\left(y-\frac{z}{L_2}\right)\left(1+\frac{y^2}{1+y^2}\right)\right\} + \ldots \right.$$
$$\left. \ldots + \frac{(\eta_0-\eta)^2}{L_2^2}\left\{1+y^2+\left(y-\frac{z}{L_2}\right)^2\left(1+\frac{\gamma^2}{4}\right)\right\}\right] \quad (30)$$

Moreover, the transition between the modes in concave and convex barriers in (22) is performed by changing the signs of $s_1$ and $s_2$ to the opposite ones. Finally, the transition to the homogeneous film, arising in a limit $\Omega = 0$, $N = 1$, $U = 1$, $\eta_0 = d$, in both cases $N^2 > 0$ and $N^2 < 0$, brings the value of group velocity in a homogeneous film $v_g = v_{go}$

$$v_{g0} = \frac{2c}{1+n_0^2} \qquad (31)$$

Thus, the expressions obtained for $\theta_\pm$ provide the general formula for 3 types of photonic barriers: convex ($N^2 > 0$), concave ($N^2 > 0$) and concave for evanescent waves ($N^2 < 0$). To calculate the values of $v_g$, one has to define the values of variables $\eta$ (5). Substitution of (3) into (5) yields for the concave profile

$$\eta(z) = \frac{L_2}{2\sqrt{1+y^2}} \ln\left(\frac{y_+}{y_-} \frac{y_- + z/L_2}{y_+ - z/L_2}\right); \quad y_\pm = \sqrt{1+y^2} \pm y$$

$$\eta_0 = \eta(d) = \frac{L_2}{\sqrt{1+y^2}} \ln\left(\frac{y_+}{y_-}\right)$$

(32)

Likewise, one finds the variable $\eta$ for the convex profile :

$$\eta(z) = \frac{L_2}{\sqrt{1-y^2}} \text{Arctg}\left[\frac{(z/L_2)\sqrt{1-y^2}}{1-yz/L_2}\right]; \quad y^2 < 1$$

$$\eta_0 = \eta(d) = \frac{L_2}{\sqrt{1-y^2}} \text{Arctg}\left[\frac{2y\sqrt{1-y^2}}{1-2y^2}\right]$$

(33)

Using (32) and (33) we will find the group velocities. To visualize the effects of heterogeneity, it is convenient to present the values of $v_g$, normalized by means of its values in homogeneous films (31) $V = v_g/v_{g0}$. These group velocities are presented in Figure 1 for three types of barriers with thickness $d$ and refractive index $n_0 = 1.8$ (at a wavelength $\lambda = 800$ nm ): transparent-convex ($d=100$ nm, $U_m = 1.25$), transparent-concave ($d=100$ nm, $U_m = 0.75$) and tunneling-concave ($d=80$ nm, $U_m = 0.75$) barriers for the different positions inside the film. Note that the value $V=1$ at the far side of the layers arise for the symmetrical profiles $U(0) = U(d) = 1$; this would not be the case for asymmetrical profile, e.g., $U(0) = 1$, $U(d) = U_m$. Thus one can see that these photonic barriers form highly dispersive films; the velocity of energy transfer through such barriers can be increased or decreased essentially due to the $\varepsilon(z)$ profile inside the barrier. One notes a remarkable similarity between the general behaviour of the group velocities for the two concave profiles, despite the fact that one of them (fig. 1,b) is transparent ($N^2 > 0$) while the other (fig 1,c) is opaque ($N^2<0$). They both differ totally from that observed in the convex case (fig. 1a).

**V - Phase shift of waves tunneling a concave dielectric barrier ( Hartman effect ).**

The group delay $t_g$ for waves, traversing the film, can be found directly from (23)

$$t_g = \frac{1}{4c}\int_0^d U\theta dz \qquad (34)$$

The values of $t_g$ for typical parameters of heterogeneous films discussed above, do not exceed the "causality limit" $t_0 = d/c$: e.g;, the ratio $t_g/t_0$ for the evanescent wave with $\lambda$ = 800 nm, tunneling through concave barrier, formed by the film with thickness d = 80 nm and $U_m$ = 0.75, found from (34) is 0.84. No phase velocity is known to be attributed to these waves. However, there is a phase shift accumulated by these waves due to tunneling. Side by side with delay time $t_g$ (34) one can define the "phase time" $t_p$, connected with the phase shift $\varphi_-$ of tunneling waves

$$t_p = \frac{\partial \varphi_-}{\partial \omega} \qquad (35)$$

This time $t_p$ is widely discussed in the literature devoted to superluminal propagation of evanescent waves through thick opaque barriers ($kd \gg 1$) /11/-/18/, where the case $d/t_p > c$ was analysed, and interpreted as "superluminal" propagation.

Unlike the traditional schemes for observation of optical tunneling, based on the inclined incidence, such as, e.g., the frustrated total internal reflection or Goos – Hanchen shift /16/, we will examine the aforesaid phase effect for the simple geometry of normal incidence of wave on the concave photonic barrier, using the results of Section IV. The quantity can be found from the complex transmission function $T_-$ (such that $E_t = E_i T_-$) for the evanescent waves (25)– (26). This function depends upon the reflection coefficient $R_-$, which can be found for these waves from the comparison of eq. (25) – (26) at the interface $z=0$, eq. (27) being taken into account :

$$R_- = \frac{\text{th}(p\eta_0)\left[1+\gamma^2/4+n_0^2 N_-^2\right]-\gamma n_0 N_-}{\text{th}(p\eta_0)\left[1-\gamma^2/4-n_0^2 N_-^2\right]+\gamma n_0 N_- + i\left[2n_0 N_- - \gamma\,\text{th}(p\eta_0)\right]} \qquad (36)$$

Substitution of (36) into (25) brings the transmission function $T_- = |T_-| \exp(i\,\varphi_-)$

$$|T_-| = \frac{2n_0 N_-}{\text{ch}(p\eta_0)}\left\{\left[\text{th}(p\eta_0)\left(1-\gamma^2/4-n_0^2 N_-^2\right)+\gamma n_0 N_-\right]^2 + \left[2n_0 N_- - \gamma\,\text{th}(p\eta_0)\right]^2\right\}^{-\frac{1}{2}} \qquad (37)$$

$$\varphi_- = \text{Arctg}\left[\frac{\text{th}(p\eta_0)\left(1-\gamma^2/4-n_0^2 N_-^2\right)+\gamma n_0 N_-}{2n_0 N_- - \gamma\,\text{th}(p\eta_0)}\right] \qquad (38)$$

The "phase time" $t_p$, calculated by means of (35) and (38), is then

$$t_p = \frac{1}{\omega}\frac{\cos^2\varphi_-}{2n_0 N_- - \gamma\,\text{th}(p\eta_0)}\cdot\left[\text{th}(p\eta_0)\left(\frac{\gamma^2}{4}+2n_0^2 u^2\right) - \left(1-\frac{\gamma^2}{4}-n_0^2 N_-^2\right)\frac{\ln\left(y_+/y_-\right)}{uN_-\,\text{ch}^2(p\eta_0)} \right.$$
$$\left. -\gamma n_0 N_-\left(1+\frac{u^2}{N_-^2}\right) + \text{tg}\,\varphi_-\left(\frac{2n_0 u^2}{N_-} - \gamma\,\text{th}(p\eta_0) - \frac{\gamma\ln\left(y_+/y_-\right)}{uN_-\,\text{ch}^2(p\eta_0)}\right)\right] \qquad (39)$$

The phase time (39) for $\lambda = 800$ nm (corresponding to the profile used for fig. 1,c, for different thicknesses $d$) $t_p \approx 0.4$; supposing, for an estimation, the velocity $v_p$ to be constant $v_p = d/t_p$, one finds the subluminal value $v_p = 0.65c$.

To examine the phase shift produced by $m$ similar adjacent films one has to generalize the transmission function $T$ (37)–(38) for a set of $m$ adjacent parallel layers, the value $m$ being arbitrary. Using the continuity conditions on each boundary between two adjacent layers, one can find the field in each layer; attributing the number $m = 1$ for the layer at the far side of the set, we will find a simple recursive relation for parameter $Q_{m+1}$ for the $(m+1)^{th}$ layer ($m \geq 1$) :

$$Q_{m+1} = \exp[-2mp\eta_0] Q_0 \qquad (40)$$

The value $Q_0$ is given in (27). Proceeding as above, we will build a generalization of $T_-$ yielding the phase of transmission function for $m$ layers :

$$\varphi_- = \text{Arctg}\left[\frac{\text{th}(mp\eta_0)\left(1 - \gamma^2/4 - n_0^2 N_-^2\right) + \gamma n_0 N_-}{2n_0 N_- - \gamma \,\text{th}(mp\eta_0)}\right] \qquad (41)$$

A plot of the phase time and phase shifts as a function of m is shown on figure 2. The phase is growing with the increase of amount of layers: thus, for $m=1$ and $m=2$ the values are 0.97 rad and 1.2 rad respectively. However, this growth is decelerating with the increase of $m$: the values for $m=5$ and $m=8$ are 1.46 rad and 1.47 rad respectively. When the amount of layers is growing so that $mp\eta_0 \gg 1$, the phase becomes independent from $m$

$$\lim \varphi_- \big|_{m \to \infty} = \text{Arctg} \left[ \frac{1 - \left(n_0 N_- - \gamma/2\right)^2}{2 n_0 N_- - \gamma} \right] \tag{42}$$

Increasing $m$ also leads to a drasstic attenuation of the power of the tunneling wave: the transmission function (27) yields the attenuation factor $\exp(-2mp\eta_0)$.

Formula (42) describes the Hartman effect for photonic barriers formed by a stack of heterogeneous dielectric layers (3). The dependence of phase shift upon the barrier's thickness being saturated, the phase time $t_p$ can become superluminal; in our example the velocity $v_p = d/t_p$ reaches the value $c$ for the wave tunneling through 10 films. The further increase of $m$ leads to superluminal values for $v_p > c$. Herein the quantity $v_p$ can reach an arbitrarily high value, and thus the problem of physical interpretation of $v_p$ remains opened.

It is worthwhile to examine the phase saturation phenomena in the reflected wave too. Presenting the reflection coefficient $R_-$ (36) in a form $R_- = |R_-|\exp(i\phi_R)$ one can find the phase

$$\varphi_R = \text{Arctg}\left[ \frac{\gamma \,\text{th}(p\eta_0) - 2n_0 N_-}{\text{th}(p\eta_0)\left[1 - \gamma^2/4 - n_0^2 N_-^2\right] + \gamma n_0 N_-} \right] \tag{43}$$

Comparison of the phases of transmitted (38) and reflected (43) waves brings the relation $\text{tg}\,\varphi_t \cdot \text{tg}\,\varphi_R = -1$; thus, these phases and the corresponding phase times (35) $(t_p)_t$ and $(t_p)_R$ are linked:

$$\varphi_t - \varphi_R = \frac{\pi}{2} \quad (t_p)_t = -(t_p)_R \tag{44}$$

Saturation of the phase of the reflected wave thus also occurs, and it is not connected with the power attenuation; the power reflection coefficient $|R_-|^2$ in a case $m \gg 1$ tends to a natural limit $|R_-|^2 \to 1$. Let us finally point out that the phase time can take negative values. To illustrate this case, let us consider the formula (39) for the saturated case - $th(mp\eta_0) \to 1$ - in the form

$$t_p = \frac{2}{\omega}\left[1+\left(\frac{\gamma}{2}-n_0 N_-\right)^2\right]^{-2}\left\{\left[\frac{\gamma^2}{2}+2n_0^2 u^2 - \gamma n_0 N_-\left(1+\frac{u^2}{N_-^2}\right)\right]\left(\frac{\gamma}{2}-n_0 N_-\right)+\frac{n_0 u^2}{N_-}-\frac{\gamma}{2}\right\} \tag{45}$$

If the normalized frequency $u$ obeys to the condition

$$\frac{\gamma}{2}-n_0 N_- = 1 \tag{46}$$

the expression (45) for $t_p$ writes

$$t_p = \frac{1}{2\omega}\left(1-\frac{n_0}{N_-}\right) \tag{47}$$

so that $t_p < 0$, if the refractive index $n_o$ is high enough: $n_0 > N_-$. This condition can be fulfilled for the solutions of eq. (46)

$$u = \frac{\sqrt{1+y^2}}{n_0}\left(-y+\sqrt{1+y^2+n_0^2}\right) \qquad (48)$$

These solutions possess the property u ≥1, and thus meet the tunneling condition. If the frequency of tunneling wave is only slightly smaller, then the cut-off frequency ($u$ -1 «1, $N_-$ «1), then the phase time (47) reaches large negative values. Thus, for the abovementioned film ($d$ = 80 nm, $y$ = 0.575, $n_0$ = 1.8, $\Omega_1$ = 2.76 $10^{15}$ rad.s$^{-1}$), $u$=1.001, and $N_-$ = 0.045. These parameters bring the tunneling wave frequency $\omega$= 2.757 $10^{15}$ rad.s$^{-1}$ ,and finally $t_p$=-7 fs.

**VI. - EM tunneling through lossy barriers (violation of the Hartman effect) .**

This chapter is devoted to the phase effects for EM waves tunneling through dissipative heterogeneous dielectrics, described by complex values of the dielectric susceptibility $\varepsilon(z) = n_0^2(1 + ig)^2 U^2(z)$, where the spatial profile $U(z)$ is given in (3) and the absorption is determined by the factor $g$. The interplay of tunneling and absorption will affect both reflected and transmitted waves. The reflection coefficient can be found by the same procedure, and the result can be presented using the following replacements in eq. (36), derived for the lossless dielectric:

$$N_- \to a_1 - ia_2; \quad a_{1,2} = \sqrt{\frac{1}{2}\left[\sqrt{(u^2-1)^2 + g^2} \pm (u^2-1)\right]}$$

$$p \to p_1 - ip_2; \quad p_{1,2} = \frac{\omega n_0}{c} a_{1,2} \qquad (49)$$

$$\text{th}(p\eta_0) = \frac{t_1(1+t_2^2) - it_2(1-t_1^2)}{1+t_1^2 t_2^2}; \quad t_1 = \text{th}(mp_1\eta_0); \quad t_2 = \text{tg}(mp_2\eta_0)$$

$m$ representing again the number of films in the stack. Making these replacements, we find the reflection coefficient:

$$\left|R_-\right| = \sqrt{\frac{K_1^2 + K_2^2}{b_1^2 + b_2^2}}; \quad \varphi_r = \text{Arctg}\left(\frac{K_2 b_1 - K_1 b_2}{K_2 b_1 + K_1 b_2}\right)$$

$$K_1 = t_1\left(1+t_2^2\right)\left(1 - \frac{\gamma^2}{4} + n_0^2 N_-^2\right) - t_2 g n_0^2 \left(1-t_1^2\right) - \gamma g n_0 a_1 \left(1 + t_1^2 + t_2^2\right)$$

$$K_2 = \gamma g n_0 a_1 \left(1 + t_1^2 + t_2^2\right) - t_2\left(1-t_1^2\right)\left(1 + \frac{\gamma^2}{4} + n_0^2 N_-^2\right) - t_1 g n_0^2 \left(1 + t_2^2\right) \tag{50}$$

$$b_1 = t_2\left(1-t_1^2\right)\left(n_0^2 g - \gamma\right) + n_0\left(1 + t_1^2 t_2^2\right)\left(2a_2 - \gamma a_1\right) + t_1\left(1+t_2^2\right)\left(1 - \frac{\gamma^2}{4} - n_0^2 N_-^2\right)$$

$$b_2 = t_1 g n_0^2 \left(1 + t_2^2\right) - \gamma t_1\left(1+t_2^2\right) + n_0\left(1 + t_1^2 + t_2^2\right)\left(2a_1 - \gamma a_2\right) - t_2\left(1+t_1^2\right)\left(1 - \frac{\gamma^2}{4} - n_0^2 N_-^2\right)$$

When the amount of films is increasing, so that th($m p_1 \eta_0$) $\to$ 1, the expressions for $K_{1,2}$ and $b_{1,2}$ simplify :

$$K_1 = t_1\left(1+t_2^2\right)\alpha; \quad \alpha = 1 + \frac{\gamma^2}{4} + n_0^2 N_-^2 - \gamma n_0 a_1$$

$$K_2 = \left(1+t_2^2\right)\beta; \quad \beta = n_0\left(\gamma a_2 - g n_0\right)$$

$$b_1 = \left(1+t_2^2\right)\chi; \quad \chi = 2 - \alpha - 2 n_0 a_2 \tag{51}$$

$$b_2 = \left(1+t_2^2\right)\delta; \quad \delta = 2 n_0 a_1 - \gamma - \beta$$

and both amplitude and phase of reflected wave tend monotonically to some constant values

$$|R_-| = \sqrt{\frac{\alpha^2 + \beta^2}{\chi^2 + \delta^2}}; \quad \varphi_r = Arctg\left(\frac{\beta\chi - \alpha\delta}{\beta\alpha + \chi\delta}\right) \qquad (52)$$

The same approach results in the generalization of transmission function (37)-(38) for the evanescent waves in a lossy dielectric. Thus, in the "saturation" limit the amplitude and phase of this function are:

$$|T_-| = \frac{2n_0\sqrt{a_1^2 + a_2^2}}{\text{ch}(mp\eta_0)\sqrt{\chi^2 + \delta^2}} \qquad (53)$$

$$\varphi_t = Arctg\left(\frac{\chi}{\delta}\right) - Arctg\left(\frac{a_2 - a_1 t_2}{a_1 + a_2 t_2}\right) \qquad (54)$$

Due to the increase of $m$ the amplitude $|T_-|$ is damping monotonically, meanwhile the behaviour of phase $\varphi_t$ is more complicated: the relation (44) between phases $\varphi_t$ and $\varphi_r$ is violated and, unlike in the lossless case, the phase of transmitted wave is not saturated. Eq. (54) indicates the oscillations of this phase between the values $(\phi_t)_+$ and $(\phi_t)_-$.

$$(\phi_t)_\pm = Arctg\left(\frac{\chi}{\delta}\right) \pm Arctg\left(\frac{a_1}{a_2}\right) \qquad (55)$$

Such $\varphi_t$ oscillations are illustrated on figure 3 for a stack of 80 nm films desribed above, between m=5 and m=20, with g=0.5 and g=0.1. They arise for an evanescent wave in a media with arbitrary small, but finite, values of the absorption parameter $g$; thus, the monotonic saturation of the phase of the transmitted wave can be attributed only to the lossless limit $g \to 0$. However, for small values of g (see the case g=0.1 on fig. 3), the period of these oscillations is slowing down, and stacks with a small nuber of films exhibit a monotonic behavior.

**Conclusions.**

In this paper, we have studied the propagation of e.m. fields in photonic barriers, formed by inhomogeneous dielectric layers, both lossless and lossy, whose $\varepsilon(z)$ profile allows to derive exact analytical solution of the wave equation. The film properties are in such a case essentially determined by the length scales (two in our case, allowing a wide variety of film types) of the distribution of the dielectric susceptibility, and exhibit a plasma- of waveguide-like behavior, characterized by a length-scale dependent critical frequency. Using different profile parameters, we can investigate situation in which the field either propagates through the film, as well as the tunneling regime for which the field in the film is presented by an evanescent wave.

The group velocity of the waves was shown to depend upon the profile of $\varepsilon(z)$ across the film, and essentially on the convex or concave nature of the susceptibility profile. They stay in all studied cases subluminal.

We also investigated the behavior of the "phase-time" of films through which the field is tunneling, on the basis of which the possibility of superluminal phase propagation was proposed earlier. Analyzing the case of a stack of identical layers, we had visualized, that this phase time showed a saturated behavior, which was accompanied by a drastic attenuation of the tunneling intensity.

**Acknowledgement.**

The authors thank Prof. T. Arecchi for his interest into this research. They also acknowledge the support of NATO grant n° PST.CLG.980334.

**Figure captions** :

Figure 1 : group velocity as a function of z in three different symmetric barriers, $n_0$ = 1.8:, (a) convex profile, d= 100 nm, $U_m$ = 1.25; (b) concave profile, d=100 nm, $U_m$ = 0.75: (c) concave profile, d=80 nm, $U_m$ = 0.75.

Figure 2 : phase of the transmission function and phase time for a stack of m films identical to that of figure 1,c)

Figure 3 : phase of the transmission function for a stack of lossy films with parameters identical to that of fig 1,c), and imaginary part of the refractive index g equal to 0.1 and 0.5.

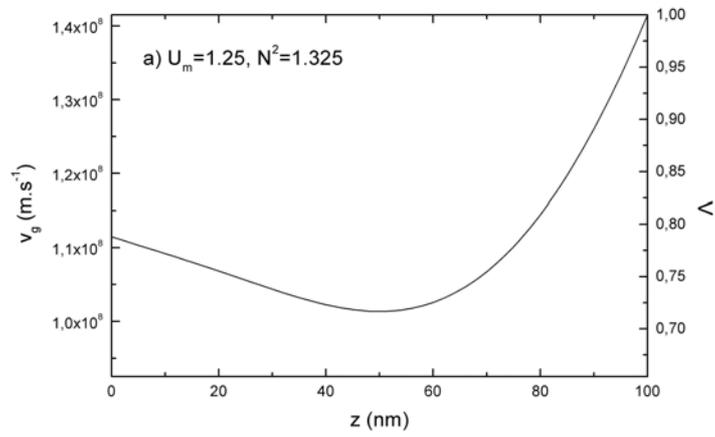

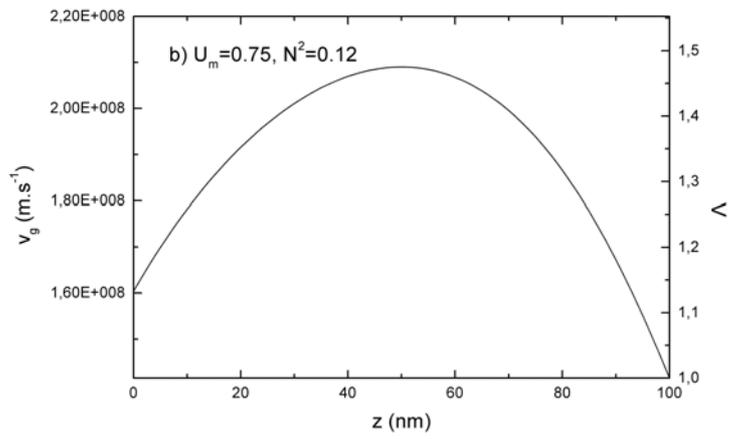

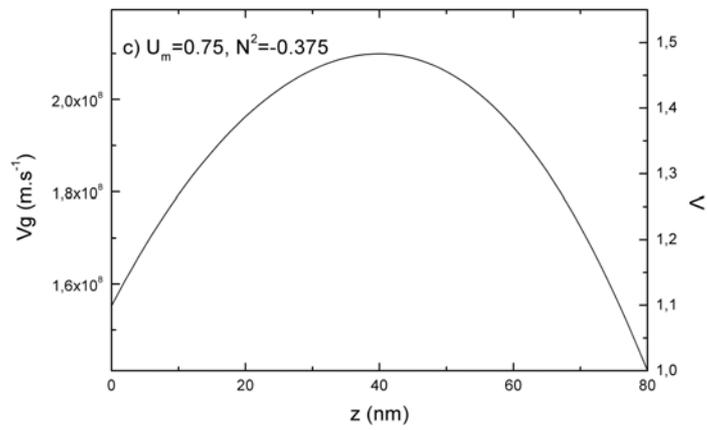

Shvartsburg & Petite, FIGURE 1

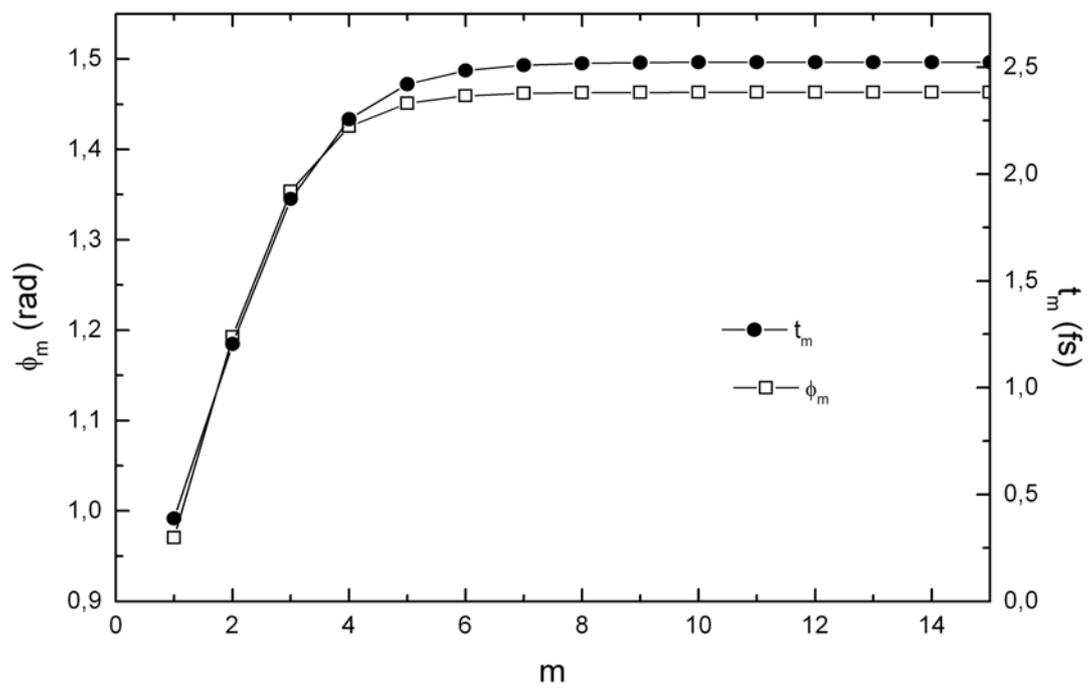

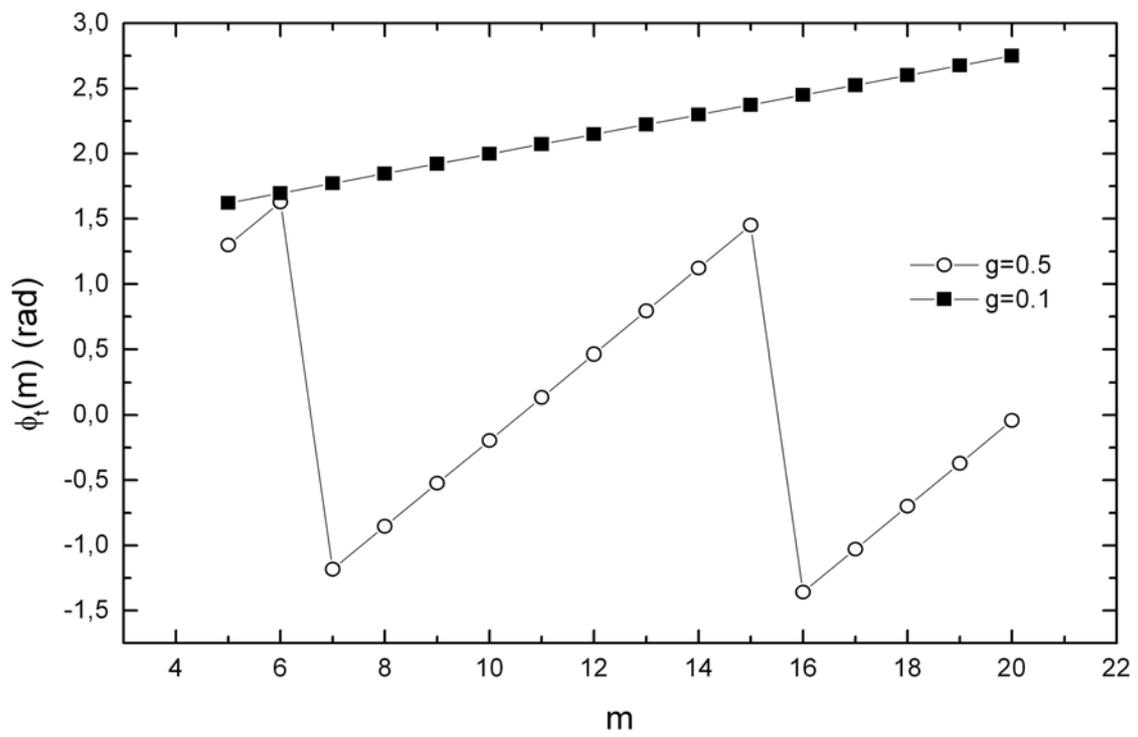

Shvartsburg & Petite, FIGURE 3